# A Conjecture Concerning Ball Lightning

by P.A. Sturrock

Stanford University

Stanford, California

**Abstract**. There is at present no theory that can explain the curious properties of ball lightning. This suggests that we may not be using the most appropriate concepts. The concept of a "parallel space" may point the way to a valid theory.

According to Barry[1], Singer[2], Stenhoff[3] and others (for a brief introduction, see [4]), some of the basic facts concerning ball lightning are as follows:

 1 . The diameter of a ball-lightning is typically in the range 10 – 50 cm. There are few reports of ball-lightnings that are very much smaller or very much larger.
 2 . The lifetime is typically in the range 1 - 5 seconds, but there are reports of longer lifetimes.
 3 . Ball-lightnings are self-luminous with a luminosity comparable to that of a few-watt lamp.
 4 . Ball-lightnings are typically described as transparent or semi-transparent rather than solid in appearance.
 5 . Ball-lightnings have varying colors, common colors being red, orange and yellow.
 6 . Ball-lightnings tend to move slowly, with speeds of order 1 meter per second, often erratically.
 7 . A ball-lightning may fade away quietly or may explode.

The phenomenon has electromagnetic characteristics:
 8 . Ball-lightnings tend to occur when and where lightning is occurring or is likely to occur.
 9 . Ball-lightnings often follow telephone lines or other electrical structures.
 10. A ball-lightning may have the appearance and odor of an electrical phenomenon, with sparkling and jittering fine structure.

11. Some witnesses have experienced electric shocks by being in contact with a metal structure that was contacted by a ball-lightning.

12. Some ball-lightnings have put a magnetic compass out of action – presumably by demagnetizing it.

13. Telephones and other electrical devices, which may be some distance away, may be put out of action at the time of a ball-lightning event.

The following facts make the phenomenon particularly intriguing:

14. A ball-lightning can move independently of the atmosphere. Jennison[5] refers to *an observation of a 20-cm ball that appeared 50 cm above the trailing edge of the wing of an aircraft in flight. It moved parallel to the wing at a speed of about 1 meter per second before being cast off at the end. The ball was not blown off despite its remarkable air speed.*

15. A ball-lightning can move through a window or even a 2-foot thick wall.[6]

16. Ball-lightnings have entered or formed within aircraft.[7] Singer mentions a case in which the pilot of an aircraft observed *a yellow-white ball approximately 45 cm in diameter enter through the windshield…*[8] When inside an aircraft, the ball-lightning is typically said to move at a steady speed of order 1 meter per second in a straight line from front to rear of the aircraft.

17. A ball-lightning may cause no damage or great damage. Some have been reported to destroy trees. Some have killed men or animals. According to analyses of some events, the energy released by a ball-lightning can be as high as 3 megajoules.

18. A ball-lightning may melt metal, for instance pitting an aircraft wing or propeller.

19. There appears to be little or no correlation between the energy released by a ball-lightning and its appearance (size, luminosity, etc.).

It would appear from their publications that Barry[1], Singer[2] and Stenhoff[3] agree with the following statement by Hill et al.: *There have been many theories advanced to explain ball lightning [but] no theory is completely satisfactory…*[9]  The purpose of this article is to suggest a concept that may lead to a satisfactory theory.

If we accept as a premise the principle of conservation of energy, leading present-day theories can be divided into two categories. In one category, energy emitted by a ball-lightning has been stored in the ball-lightning itself. In the other category, energy emitted by a ball-lightning is fed into the ball-lightning as an electrical current or as electromagnetic waves such as microwaves.

Barry, Singer and Stenhoff consider a number of stored-energy models but find none satisfactory. The fact that some ball-lightnings can move independently of the atmosphere is a problem for all such models. (See, for example, item 14.) Another general problem, noted by Singer, is that typically there is no decrease of size or brightness or change of color during the lifetime of a ball-lightning.[10] Finkelstein and Rubenstein examined the implications of the virial theorem for plasmoid models, and found that it sets too low a limit on the energy that can be stored in such a structure.[11] The virial theorem holds not only for a nonrelativistic plasma configuration but also for a relativistic plasma configuration such as the *spherical plasma bubble* model recently proposed by Wu.[12]

A major problem with injected-current and injected microwaves proposals is the difficulty of understanding how an electrical current or electromagnetic waves could penetrate the metal shell of an aircraft.

Since the two current categories of theory are widely considered inadequate for explaining the properties of ball lightning, it seems there is nothing to be lost in looking for a third category.

We here argue as follows:
( a ) Since there is no known way for the required energy to be stored in the ball-lightning, there must be a reservoir of energy remote from the ball-lightning (presumably related to the electrical energy responsible for lightning).

( b ) Since the reservoir is remote from the ball-lightning, there must be some way to transfer energy from the reservoir to the ball-lightning. We therefore conceive of a *duct* that connects the reservoir to the ball-lightning.

( c ) A ball-lightning may now be regarded as a *port* through which energy in the duct can be released into the atmosphere.

Concerning the duct, we require that, in addition to its electromagnetic properties or capabilities,
( a ) its motion is not restricted by the atmosphere;
( b ) it can penetrate a wall or window without causing any damage;
( c ) it can penetrate a metal structure such as an aircraft fuselage; and
( d ) it is invisible.

These characteristics are suggestive of a modification of our familiar *overt* space, which we can think of as a different but parallel *covert* space. The transition from the overt space to the covert space may be an on-off proposition or a matter of degree.

These thoughts suggest the following hypothesis:
*A ball-lightning is a port connecting our overt space to a covert space with similar but not identical properties.*

This model seems to be compatible with items 1 through 7. Just as the appearance of a household electrical outlet bears no relation to the current being drawn from the outlet, this model can explain why the size, luminosity and other manifest properties of the ball-lightning seem to bear no relation to the energy released by the ball-lightning.

This model seems also to be compatible with items 8 through 13, since in this model a ball-lightning is coupled to a remote reservoir of electromagnetic energy. We note in particular that this model can accommodate item (13); the reservoir may be far from the ball-lightning so that the duct may have influences far from the ball-lightning. The sudden eruption of a duct to form a ball-lightning may trigger a disturbance throughout the duct that results in electromagnetic events remote from the ball-lightning (reminiscent of an Alfven wave traveling along a magnetic flux tube).

Is there any evidence for a duct in ball lightning events? Singer mentions two cases in which a bright *ray* or *line of fire* extends from a ball-lightning.[13] These rays may be manifestations of the hypothesized ducts.

Is there any evidence that the interior of a ball-lightning has unusual properties? Singer mentions a case in which two witnesses encountered a large bright ball 4 m in diameter.[14] *The ball sank through the telegraph wires, which glowed, and then enveloped the couple. They stood in a thick white sea of light in which the sensations of odor or heat were absent. There was no breeze from the motion of the ball, and they could not feel the outside wind. They could see only the pebbles of the road.*

Are there any other phenomena that have points of similarity with this concept of ball lightning? We have recently drawn attention to the phenomenon known as *Mobile Luminous Objects* (MLOs) that form in superconducting cavities at very low temperatures in response to strong radiofrequency electromagnetic fields.[4,15] These seem to resemble ball-lightnings but are much smaller, with diameters of order 1 millimeter. An MLO may be an exit port rather than an entry port: electromagnetic energy from the RF field may pass through a mini-ball-lightning (an MLO) to inject energy into a reservoir. After the RF field is turned off, the reservoir may return some or all of the energy in the reservoir for a short time via the same mini-ball-lightning.

We may also ask whether there are any events in which some kind of object has electromagnetic properties but is invisible. We draw attention to highly reputable reports from an aircraft traffic control site of an object that was tracked by radar but was invisible.[16]

Is there any way to examine this model experimentally? The death of Georg Wilhelm Richmann was caused by a ball-lightning. Following a strike on a lightning conductor, Richmann's friend Sokolov (who was present in Richmann's laboratory) saw a fireball leave the ungrounded end of the conductor in the laboratory and float through the air to strike Richmann's forehead. There was a sound like small cannon: Sokolov lost consciousness and Richmann was killed.[11]

One could try to duplicate this event (hopefully without the fatality) by imposing a very high voltage (supplying a very high current) on a conductor penetrating a protective metallic chamber. Such events have occurred by accident in connection with the switching of submarine batteries.[17] There have been attempts to initiate similar events by triggering a discharge by the rocket-and-wire technique.[9] Another avenue of research would be to pursue the investigation of MLOs.[15]

I thank David Fryberger, Timothy Grove, Hal Puthoff, Jeff Scargle, Daniel Sheehan and Martin Uman for helpful comments and advice.

**References**

[1] Barry, J.D., *Ball Lightning and Bead Lightning, Extreme Forms of Atmospheric Electricity* (Plenum Press New York and London, 1981).

[2] Singer, S., *The Nature* of *Ball Lightning* (Plenum Press New York and London, 1971).

[3] Stenhoff, M., *Ball Lightning: An Unsolved Problem in Atmospheric Physics* (Kluwer Academic, New York, 1999).

[4] Sturrock, P.A., *Late Night Thoughts About Science*, 5 (Exoscience, Palo Alto, 2015).

[5] Jennison, R.C., *Can Ball Lightning Exist in a Vacuum ?* , Nature 245, 95 (1969).

[6] Ref. 2, p. 37.

[7] Jennison, R.C., *Ball Lightning*, Nature 224, 895. (1969).

[8] Ref. 2, p. 40.

[9] Hill, J.D., et al., *Attempts to Create Ball Lightning with Triggered Lightning*, J. Atmos. and Solar-Terr. Phys. 72, 913 (2010).

[10] Ref. 2, p. 93.

[11] Finkelstein, D., & Rubenstein, J., *Ball Lightning*, Phys. Rev. 135A, 390. (1964).

[12] Wu, H.-C., *Relativistic-microwave theory of ball lightning*, Nature, Scientific Reports 6:28263 (2016).

[13] Ref. 2, pp. 29, 39.

[14] Ref. 2, p. 45.

[15] Anthony, P.L., et al., *Experimental Studies of Light Emission Phenomena in Superconducting RF Cavities*, Nuclear Instruments and Methods in Physics A 612, 1 (2009).

[16] Condon, E.U., & Gillmor, D.S., *Final Report of the Scientific Study of Unidentified Flying Objects* (Bantam, New York, 1969), p. 310.

[17] Silberg, P.A., *Ball Lightning and Plasmoids*, J.G.R. 67, 4941 (1962).